\documentclass{article} 
\begin{document} 
\begin{center} 
\title{Symmetry limitations on quantum mechanical observers, and conjectured link with string theory} 
\author{M.Dance}
\maketitle 
\end{center} 

\begin{abstract} 
This paper tentatively conjectures a possible physical picture that may help explain links between quantum field theories and string theories.  A correspondence might occur if the stringy parameters $\tau $ and $\sigma_i $ are interpreted as representing particular types of observer capabilities. Observer limitations could be imposed by symmetries of relevant quantum mechanical observer states. It is possible that this postulate may in future be able to go a small way towards understanding a physical basis for some of the dualities that have been found in string/field theory. 
\end{abstract} 

\section{Introduction}

String theory dualities appear to point to the existence of a unifying candidate theory for quantum gravity, dubbed M-theory.  The fundamental physical basis of M theory is unknown. 

Significant progress has been made in the last decade on noncommutative geometry theories and links with string theory.  Noncommutative geometry is linked to the inclusion in of observers that obey quantum mechanics into special relativity. It seems possible that further consideration of quantum mechanical observers may be useful to increase our understanding of nature. 

Dirac noted~\cite{dirac} that observation is inherently problematic in quantum mechanics, while being critical in relativity. Heisenberg noted that his uncertainty principle applies to observers as well as to observed systems~\cite{heisenberg}. He postulated that any practical effect of observer indeterminacy could be eliminated by allowing the observer's mass to approach infinity. (However, gravity would then become important.) A long time ago, Feynman and Hibbs~\cite{feynman&hibbs} noted that the usual separation of observer and observed in quantum mechanics should not be necessary. However, they viewed this as a purely philosophical issue, analysis of which was unnecessary for the further development of physics. 

Nowadays, minimum observable distance and time intervals are well known for observers that obey quantum mechanics. These intervals have been linked with non-commutative geometries based on the Poincare group. In these geometries, it is as if the observer perceives a spacetime with noncommuting coordinates, with modified calculus based on the Moyal star product in place of ordinary multiplication. The concept of an inertial reference frame becomes a quantum superposition of reference frames. Non-commutative geometry theories formulate special relativity in a way that is consistent with quantum mechanics. 

Physically, non-commutative geometries incorporate quantum mechanical uncertainty in the position and momentum of an observer's centre of mass. 

Beyond this, it seems possible that a richer theory remains to be uncovered, in which one takes into account further properties of a quantum mechanical observer's states. Bander~\cite{bander} has recently shown that non-commutative geometry is an effective low-energy theory of systems coupled to an auxiliary system. One might interpret the auxiliary system of \cite{bander} as an observer.

This paper explores the effect of symmetries of quantum mechanical observer states. It is possible that such symmetries can restrict the information that is accessible to observers, in interesting ways.

\section{The observer}

Let us begin by considering which part of any measuring apparatus one views as "the observer". In  my view, all observations at the quantum level are carried out by quantum mechanical observers (assuming the framework of quantum theory). Examples are a single atom or molecule, e.g. an atom in a plasma, or e.g. a retinal molecule in the retina. More generally, an observer could be any suitable quantum field. In practice, the initial observer is generally an electron, in an atom or molecule.

For example, the retina detects a single photon by altering the conformation of a single covalent bond within a single retinal molecule. The conformation change is an inherently quantum mechanical event. Later stages of visual processing amplify and filter the initial quantum mechanical observation event. This paper will not consider these later stages, but only the initial quantum mechanical interaction of the observed system with the observer.

A useful approach to formulating quantum field theories consistent with quantum mechanical observers may be to simply add one or more observer terms to field theory actions. That is the approach adopted in this paper (although if time permitted, this author would ideally wish to reconsider the foundations of Lagrangian mechanics before doing so). One can then consider the states of the observed system and of the observer. From the "initial" and "final" observer states, one could determine what the observer observes. Observer states would be included explicitly in S-matrix calculations. 

Observer terms need not render theories unduly complex. A quantum observer can be a simple entity, such as an electron in an H atom. Furthermore, it may be that observer terms need take into account only selected observer features, to give useful working approximations.

This paper considers a quantum mechanical observer and observed system as seen by a perfect classical observer, i.e. in conventional spacetime. It remains to be seen whether this approach is valid, but it may be a useful perturbative approach.

\section{Observers in action}

To incorporate observer dynamics, we now add a simple observer term to field theory Lagrangian densities. In flat spacetime, a very simple non-relativistic term could represent the observer's kinetic energy in the form:-  

\begin{equation}
L^{EK}_{obs} = \frac{1}{2}m \eta_{ij}\frac{dX^i}{dT}\frac{dX^j}{dT}
		= \frac{1}{2}m \kappa^{\alpha \beta } \eta_{\mu \nu } 
		\frac{\partial{X^\mu}}{\partial{x^\alpha}}  
		\frac{\partial{X^\nu}}{\partial{x^\beta}}
\end{equation}
where $m$ is the mass of the observer state (e.g. an electron), $x^{\alpha }$ and $x^{\beta }$ are the coordinates ($t, x, y, z,..$) internal to the observer, and ${X^{\mu }}$ are fields representing the observer's position in classical coordinates.  One might think of these classical coordinates as the coordinates of a second, classical, observer. (Perhaps this scheme would need to be iterated indefinitely to be completely self-consistent.) $\kappa $ is a factor which depends on the transformation between the $x^{\alpha } $ and $X^{\mu } $ coordinates. $\kappa $ may depend on $X$.  The observer term above is very simple, but it may be a start.

At first sight, it might seem strange to have a term for the observer kinetic energy, expressed in the observer's own internal coordinates.  But quantum mechanics does strange things. The observer itself cannot know where it is with complete precision at any one time, relative to where it may have been at any other internal time. Its knowledge (if it can be called that) of itself is limited at the very least by the quantum mechanical minimum observable distance and time interval.

\section{String theory}

The simplest string action is the area of the world-sheet of a string propagating through flat space-time, multiplied by a constant tension factor. In conformal gauge, the simple action is:-
\begin{equation}
S_{string} = 	-\frac{T}{2} \int d\tau d\sigma
		\eta^{\alpha \beta } \eta_{\mu \nu } 
		\frac{\partial{X^\mu}}{\partial{x^\alpha}}
		\frac{\partial{X^\nu}}{\partial{x^\beta}},
\end{equation}
where $X^{\mu }$ are the space-time coordinates of a point on the string, and
$x^{\alpha }$, $x^{\beta}$ are members of ${\tau, \sigma}$.

\section{Observer and string terms compared}

The two terms above look somewhat similar. It is interesting to consider whether they could possibly be equivalent in some circumstances. Is it possible that a simple term for the centre-of-mass motion of a nonrelativistic observer could be a basis for string theory, if further assumptions or conditions are imposed? The observer term above is very simple, simplistic in fact, but then perhaps string theories are also each a cut-down version of reality.

Equating (in some loose sense) the factors in front of the derivatives would mean that the string tension would be related to the observer mass, i.e. perhaps to the mass of the electron. This relationship may seem unlikely, but perhaps other work will modify it, or shed further light on it. The main focus of this paper is on possibly useful ideas for the derivative terms.

An essential qualitative difference between the above Lagrangians is that string theory uses the worldsheet parameters ($\tau, \sigma_i$ ) instead of internal observer space-time coordinates $(t,x,y,z,...)$.  Can a physical correspondence be made between $(t,x,y,z,...)$ and $(\tau, \sigma_i)$?  This paper conjectures that a correspondence may arise due to limitations in observer capabilities.

\section{Observer symmetries and limitations on observation}

For a quantum mechanical observer O to observe an incoming particle/field P, the observer must undergo a transition between two of its internal quantum states. To estimate the likelihood of observation and the information which the observer can extract, we must consider the initial and final observer states, and the initial and final states of the incoming particle/field. The probability of observation is related to the amplitude of the OP interaction between the initial and final states. 

Observer symmetry is important. It impacts on transition amplitudes and on extractable information. 

I postulate here that an observer will not be able to extract information about parameters with respect to which the observer's initial and final states are completely symmetric in its own internal coordinate system, or with respect to which the observer's interaction matrix element with the observed system is zero. Literally, symmetry means "same measure". For example, one symmetry is Poincar\'e invariance. (Noncommutative geometry introduces complications here, but one can redefine operators accordingly.) A result of Poincar\'e symmetry is that no observer can detect absolute position or velocity, or absolute direction in space-time. This is the case even at the classical level.

At the quantum level, a common observer symmetry is rotational symmetry. For example, consider an H atom whose initial and final states both possess a spherically symmetric $1s$ electronic state. To this H atom observer, the concept of the angular polar coordinates 
$\theta $ and $\phi $ of an incoming particle's momentum may be meaningless. This is because the H atom has no internal reference axis on which to project its change in centre-of-mass momentum. (However another observer, e.g. our classical observer watching the H atom, may be able to detect $\phi $ and $\theta $ components of the H atom's momentum change in the classical observer's $X^{\mu}$ coordinates.)

If the H atom begins and ends an observation with its electron in (say) a $2p_z$ state, then the H atom has an internal $z$ axis. Let us suppose that it could thereby detect the magnitude of a change in its total $z$ momentum component. On the other hand, the concept of the polar coordinate $\phi $ may be meaningless to this observer.

The more symmetries an observer possesses, the less the observer may be able to detect.

\section{1D-observers and 2D-observers}

Let us call an observer that can detect one spatial coordinate a "1D-observer".

Let us firstly consider situations in which the initial and final observer states have the same symmetry, in the observer's coordinates.

If the observer has a single axis of symmetry, we can define that axis to be the observer's $z$ axis. In certain contexts, this observer may be able to detect the polar coordinate $\theta $ but cannot detect $\phi $. Let us call such an observer a $\theta $-observer.

$\theta$ exists in the interval $[0, \pi ]$. This interval corresponds with the conventional interval $[0, \pi ]$ for $\sigma$ in open string theories. It is tempting to equate $\sigma $ in open string theory with the $\theta$ coordinate of a $\theta $-observer.

If the observer has two orthogonal axes of rotational symmetry, we can define these axes to be the observer's $x$ and $y$ axes. In certain contexts, this observer may be able to detect the polar coordinate $\phi $, but cannot detect $\theta $. Let us call this observer a $\phi $-observer.

$\phi $ is in the interval $[0, 2\pi ]$, where $\phi = 0$ and $\phi = 2\pi $ represent the same space-time point. This resembles closed string theory, for which $\sigma $ is in $[0,\pi ]$, and $\sigma = 0$ and $\sigma = \pi $ represent the same point on the string. In fact, the $\sigma $ interval limits are immaterial in string theory. It is only necessary that the interval limits be different numbers. The values $0$ and $\pi $ are chosen for convenience. So it appears that the closed string $\sigma$ interval could be redefined as $[0, 2\pi ]$ to match the $\phi $ interval. It is tempting to equate $2\sigma $ in closed string theory with the $\phi $ coordinate of a $\phi $-observer.

Now let us consider observations for which the initial and final observer states possess different symmetries. For example, some observations may turn a $\theta $-observer into a $\phi $-observer. Other observations might cause an observer's $z$ axis to rotate. What can these observers observe? Perhaps these observers may be able to simultaneously detect both $\theta $ and $\phi $, or two different $\phi$  angles, respectively. It is tempting to make an analogy with branes. Alternatively, this picture might correspond to an interaction of open and closed strings/branes in a string theory perspective.

I postulate that a simple quantum system (such as a single atom) cannot generally measure the radial coordinate $r$ in its coordinate system. Even in situations where the quantum mechanical minimum measurable distance interval is much smaller than $r$, quantum mechanical transition amplitudes do not tend to pick out any particular $r$ interval in a quantum mechanical observer's internal coordinates.

In 3 space dimensions, I postulate that the simple quantum mechanical observer generally can observe at most $\theta $ and $\phi $. I will call quantum observers that can measure both "2D-observers."

\section{Conjectured correspondence}

Can the internal observer coordinates $(t,x,y,z,....)$ correspond in some sense with $(\tau ,\sigma_1,\sigma_2,...)$ ?

A 1D-observer would perceive an effective Lagrangian that is the full Lagrangian integrated over its invisible variables.  This would leave its $t$ variable and an internal angular coordinate, $\theta $ or $\phi $. Without undertaking mathematical working, I postulate that the effective Lagrangian would correspond to a string theory Lagrangian, with $t = \tau $ and $\theta = \sigma $ (or $\phi = \sigma $). I would think that the algebra would be the same.
Similarly for the algebra seen by 2D-observers: $\tau $ would correspond to $t$ and the observable angular variables would correspond to the $\sigma_i$ of string theory. 

If a change in observer symmetries can result in a change in its observations from closed to open strings and vice versa, one senses that there may be potential for dualities between closed and open strings to emerge. One also senses a potential to relate gauge theories to string theories in this physical picture, as a string term would relate to an observer limited by symmetry and gauge terms are also related to considerations of an observer's perspective.

\section{Holographic principle}

A previous paper~\cite{dance} discussed the holographic principle in the context of postulated limitations on observer capabilities. The holographic principle asserts that the maximum number of observable degrees of freedom in a volume should be proportional to the surface area of that volume. The principle gives rise to the idea of dimensional reduction. That is, observers sitting on the surface see a quantum field theory for the surface that corresponds with a (different) quantum field theory governing the volume. 
 
As above, it was postulated in~\cite{dance} that a simple quantum mechanical observer such as an atom cannot by itself measure the radial coordinate $r$ of any event in its internal (or other) coordinates. It was postulated that it requires an observer that is an open thermodynamic system to measure $r$. The observer generates entropy during its observation of $r$, and the entropy generated could effectively wipe out a dimension, in terms of the information that the observer can know at any one time about the observed system. The argument also invoked the minimum observable distance for quantum mechanical observation.

\section{Conclusion}	

This paper has suggested adding a simple observer term to conventional field theory actions. This paper also suggests that internal observer symmetries may effectively eliminate some spatial coordinates, internal to the observer, from observation. Such theories may shed some small amount of light on string theories and their dualities, e.g. if the parameters $\tau $ and $\sigma_i$ can be interpreted as representing particular types of observer capabilities.

\end{document}